\documentclass[10pt,prl,aps,showpacs,floatfix,twocolumn,unsortedaddress]{revtex4-1}

\usepackage{graphicx, graphics, bm, psfrag, amsmath, amssymb, epsfig, grffile, subfigure}

\newcommand{\eps}{\epsilon}
\newcommand{\beq}    {\begin{equation}}
\newcommand{\enq}    {\end{equation}}
\newcommand{\ceq}[1] {(\ref{#1})}
\newcommand{\kk}{{\bf k}}

\newcommand{\qq}{{\bf q}}

\newcommand{\df}     {\equiv}

\newcommand{\dfun}   {$\epsilon(\qq,\omega)$ }
\begin{document}

\title{Screening and collective modes in gapped bilayer graphene}

\author{C. Triola, E. Rossi}
\affiliation{
             Department of Physics, College of William and Mary, Williamsburg, VA 23187, USA
            }
\date{\today}


\begin{abstract}
We study the static and dynamic screening of gapped AB-stacked bilayer graphene.
Unlike previous works we use the full 4-band model instead of the simplified
2-band model. We find that there are important qualitative differences
between the dielectric screening function obtained using the simplified 2-band
model and the  4-band model. In particular, within the 4-band model, in the presence of a band gap,
the static screening
exhibits Kohn anomalies that are absent within the simplified 2-band model.
Moreover, using the 4-band model, we examine the
effect of trigonal warping
on the
screening properties of bilayer graphene. We also find that
the plasmon modes have a qualitatively different character in the 4-band model 
compared to the ones obtained using the simplified 2-band model.
%
\end{abstract}


\maketitle


Bilayer graphene
\cite{novoselov2006}
has many unique electronic properties that make it an extremely interesting
system. AB-stacked bilayer graphene (BLG) is formed by two Bernal stacked layers of graphene
\cite{novoselov2004,neto2009,dassarma2011}.
When placed on an insulating substrate the electrons in BLG
form an ideal two-dimensional electron gas (2DEG)
with a very high room temperature mobility,
in particular when boron nitride is used as a substrate
\cite{young2010,*dean2010,*ponomarenko2011}.
In pristine BLG the conduction and valence bands
touch at points, charge neutrality points (CNPs),
at the corners of the Brillouin zone.
At very low energy around these points the bands are
approximately parabolic.
However, by applying a perpendicular electric field
a band-gap ($\Delta$) can be opened, and tuned
\cite{castro2007,*min2007,*oostinga2007,*taychatanapat2010,*zou2010,*yan2010}.
Moreover, recent experiments
\cite{weitz2010,*mayorov2011,*freitag2011,*velasco2011}
have shown strong evidence that at low temperatures
and dopings the electrons in BLG might be in
a spontaneously broken symmetry state
\cite{min2008b,*vafek2010,*lemonik2010,*nandkishore2010,*nandkishore2010c,*zhang2010,*gorbar2012}.
All these facts make BLG an extremely interesting system
both from a fundamental point of view and for its possible
technological applications.
As a consequence the accurate knowledge of the electronic properties
of BLG is of great interest.

\begin{figure}[htb]
 \begin{center}
  \centering
  \includegraphics[width=8.5cm]{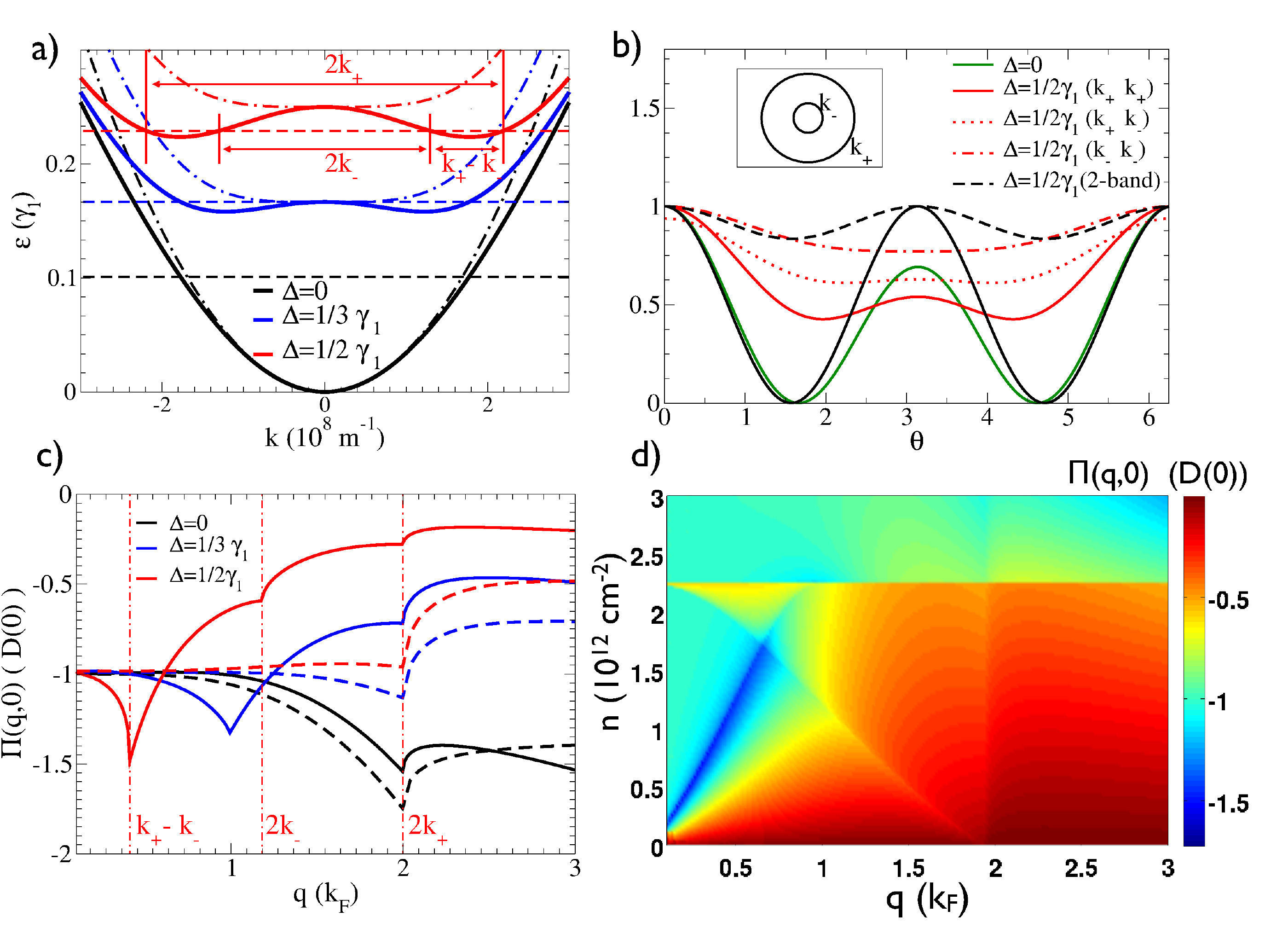} 
  \caption{
           (Color online)
           (a) Lowest conduction band for $\Delta=\gamma_1/2, \ \gamma_1/3, \ 0$.
               The solid (dash-dot) curves are obtained using the 4-band (2-band) model.
               The horizontal dashed lines indicate the 4-band Fermi energy for doping $n=10^{12}\text{cm}^{-2}$
               for  $\Delta=\gamma_1/2, \ \gamma_1/3,\ 0$ from top to bottom.
           (b) Chirality factors, $|(U^\dagger_{\textbf{k}}U_{\textbf{k}'})_{\lambda\lambda}|^2$, evaluated for $|\textbf{k}|=|\textbf{k}'|=k_F$
               for the 2-band model at $\Delta=0, \ \gamma_1/2$, denoted by the black solid and dashed lines respectively, and the 4-band model for $\Delta=\gamma_1/2, \ 0$.
               $\theta$ is the angle between $\kk$ and $\kk'$.
               For the case $\Delta=\gamma_1/2$ in the 4-band model there are three possible intraband overlap scenarios we can consider:
               (i) $\textbf{k}$ and $\textbf{k}'$ both lie on the Fermi surface at wavevector
               $k_{F+}$, (ii) $\textbf{k}$ and $\textbf{k}'$ both lie on the Fermi surface at wavevector $k_{F-}$,
               (iii) $\textbf{k}$ lies on the Fermi surface at wavevector $k_{F+}$ while $\textbf{k}'$ lies on the fermi surface at $k_{F-}$.
           (c) $\Pi(\qq,0)$ for $n=10^{12}\text{cm}^{-2}$ without trigonal warping.
               Solid (dashed) curves are the  results obtained using the 4-band (2-band) model.
           (d) Contour plot of polarizability, $\Pi(q,n,\omega=0)$, as a function of $q$ and doping $n$ for $\Delta=1/2\gamma_1$.
          }
  \label{fig:gapped}
 \end{center}
\end{figure}

One of the most important physical quantities for characterizing
the electronic properties of a system is the dielectric function
$\epsilon(\qq,\omega)$. This quantity determines the effective,
screened, Coulomb interaction among the electrons in the system
and is therefore essential for the calculation of all the electronic
properties. There is strong evidence that in most BLG
samples charge impurities
close to the surface of the substrate, or placed between the substrate
and the BLG layer, are
the dominant source of scattering
\cite{dassarma2011}.
In this situation the knowledge of the static dielectric function,
$\epsilon(\qq,\omega=0)$, is essential to calculate the dc conductivity.
Moreover, in the case of magnetic adatoms placed on BLG, the
static polarizability determines the effective
Ruderman-Kittel-Kasuya-Yoshida (RKKY) interaction between the
magnetic adatoms.
The dynamic dielectric function determines the optical properties
of the system and the collective electronic modes, plasmons.
It is therefore evident that the knowledge of the correct form
of \dfun is necessary to characterize the electronic properties of BLG.
Previous works
\cite{wang2007,*hwang2008a,*borghi2009,*gamayun2011,*yuan2011,sensarma2010}
have studied the case of gapless BLG (and gapless single and multilayer systems \cite{dassarma2009,*hwang2009b,*min2012}).
In the presence of a gap some of the symmetries that simplify the calculation
of the response functions in gapless BLG disappear. In part for this reason
the only results available \cite{wang2010} for the dielectric function in gapped BLG
were obtained using a  simplified
effective low energy 2-band model
\cite{mccann2006b,nilsson2008}.
This model neglects
features of the band structure of BLG that can strongly affect the response function,
especially when $\Delta\neq 0$.
In particular, 
it neglects the fact that
%
in the presence of
a band-gap the bands, at low-energy, acquire a characteristic ``sombrero''
shape \cite{mccann2006b}, see Fig. \ref{fig:gapped}~(a).
To describe these effects it is necessary to use 
a refined 2-band model \cite{mccann2006b,nilsson2008}
or the full 4-band model.
In this work  
we obtain \dfun
for gapped AB-stacked bilayer graphene
using the full 4-band model
and the random phase approximation (RPA).
Some of the qualitative differences for \dfun between the full
4-band model and the simplified 2-band model 
%
can be recovered using the refined 2-band model. 
However, there are features
of \dfun (especially at large $k$, $\omega$ and $n$)
obtained using the full 4-band model that are qualitatively different
from \dfun obtained using either the simplified or the refined 2-band model.
%
%
In the remainder, unless specified, by 2-band model we refer to the simplified one.
%

The 4-band continuum model Hamiltonian for BLG is $H_0=-\sum_{\textbf{k}}\Psi^\dagger_\textbf{k}h(\textbf{k})\Psi_\textbf{k}$
%
%
where
$\Psi^\dagger_\textbf{k}$ ($\Psi_\textbf{k}$) is the 4-component creation (annihilation) operator
$\Psi^\dagger_\textbf{k}=(a^\dagger_{\textbf{k},1},b^\dagger_{\textbf{k},1},a^\dagger_{\textbf{k},2},b^\dagger_{\textbf{k},2})$
($\Psi_\textbf{k}=(a_{\textbf{k},1},b_{\textbf{k},1},a_{\textbf{k},2},b_{\textbf{k},2})$)
with $a^\dagger_{\textbf{k},i}$ ($a_{\textbf{k},i}$), $b^\dagger_{\textbf{k},i}$ ($b_{\textbf{k},i}$)
the creation (annihilation) operator for an electron  with wavevector $\textbf{k}$ in layer $i$ on sublattice A and B respectively, and
$h(k)$ is the matrix
\begin{align}
 h(\textbf{k})&=\dfrac{\Delta}{2}\tau_z+\hbar
 v_F(k_x\sigma_x+k_y\sigma_y) -
 \dfrac{\gamma_1}{2}(\sigma_x\tau_x+\sigma_y\tau_y) \nonumber \\
 &+ \dfrac{3}{2}\gamma_3a\left[k_x(\sigma_x\tau_x-\sigma_y\tau_y)-k_y(\sigma_x\tau_y+\sigma_y\tau_x)\right].
 \label{eq:h}
\end{align}
In Eq.~\ceq{eq:h} $\sigma$'s, $\tau$'s are 2x2 Pauli matrices representing
the sublattice and layer degrees freedom respectively, $v_F$
is the Fermi velocity at the Dirac point of a single graphene layer,
$\gamma_1$, $\gamma_3$ are the interlayer hopping parameters \cite{note01},
%
$a=1.42\mathring{\text{A}}$ is the in plane carbon-carbon distance,
and $\Delta$ is the band gap at $k=0$. 
$\gamma_3\neq 0$ induces trigonal warping.
For concreteness, we assume
$v_F=10^6$m/s, $\gamma_1=0.35$eV, and $\gamma_3=(3/4)\gamma_1=0.26$eV,
however the main features of our results do not depend on the precise
values chosen for these parameters.

The Coulomb interactions are described by the Hamiltonian
$H_i=(1/2A)\sum_\qq[V_+(q)\hat\rho_\qq\hat\rho_{-\qq}+V_-(q)\hat d_{\qq}\hat d_{-\qq}]$,
where $A$ is the sample area,
$\hat\rho_\qq$ ($\hat d_{\qq}$ ) the operator for the sum (difference) of the densities $\hat\rho_{\qq,i}$ in the two layers,
$V_{\pm}(q)=(V_S(q)\pm V_D(q))/2$ with $V_S(q)=2\pi e^2/(\epsilon q)$ the Coulomb interaction between electrons
in the same layer and $V_D=2\pi e^2(e^{-qd})/(\epsilon q)$ the Coulomb interaction between electrons in different layers,
$d=3.35\text{\AA}$ the distance between the two layers, and $\epsilon$ the background dielectric constant. 
We assume $\alpha\df e^2/\epsilon\hbar v_F=0.5$ and 
temperature $T=0$.
As long as $q\ll 1/d$
the dielectric function that
enters the calculation of most of the electronic quantities is the one associated with the sum of the densities in the two
layers, $\eps(\qq,\omega)\df\eps(\qq,\omega)_{\rho\rho}$.
Within the random phase approximation
\begin{equation}
 \epsilon(\textbf{q},\omega)_{\rho\rho}=1-V_+(\qq) \Pi(\textbf{q},\omega)_{\rho\rho}
 \label{eq:diel-f}
\end{equation}
where
\begin{align}
 \Pi(\textbf{q},\omega)_{\rho\rho}&=g\sum_{\lambda,\lambda'}\int\dfrac{d\textbf{k}}{(2\pi)^2}
 \dfrac{n_{\lambda,\textbf{k}}-n_{\lambda',\textbf{k}+\textbf{q}}}
 {\hbar\omega+\epsilon_{\lambda,\textbf{k}}-\epsilon_{\lambda',\textbf{k}+\textbf{q}}+i\eta} \nonumber \\
 &\times |(U_{\textbf{k}}^{\dagger}U_{\textbf{k}+\textbf{q}})_{\lambda,\lambda'}|^2
 \label{eq:Pi}
\end{align}
is the polarizability (in the remainder of this paper the subscript $\rho\rho$ will be understood).
In Eq. \ceq{eq:Pi} $g=g_sg_v=4$ is the total spin ($g_s$) and valley ($g_v$) degeneracy,
$\lambda$, $\lambda'$ are the band indices, $n_{\lambda,\textbf{k}}$ ($\epsilon_{\lambda,\textbf{k}}$) is
the Fermi-Dirac distribution function (energy) for a particle in band $\lambda$ with
wavevector $\textbf{k}$,
and $U_{\textbf{k}}$ is the unitary matrix
that diagonalizes the Hamiltonian $H_0$.
From Fig.~\ref{fig:gapped}~(b) we see that the intraband wave-function overlap
$|(U_{\textbf{k}}^{\dagger}U_{\textbf{k}+\textbf{q}})_{\lambda,\lambda}|^2$
for the 4-band model is quite different from the one for the 2-band model, 
especially when $\Delta\neq 0$.
Below we present our results, obtained using Eq.~\ceq{eq:diel-f}-\ceq{eq:Pi} and 
evaluating the integral on the r.h.s of
Eq~\ceq{eq:Pi} numerically using an adaptive integration routine
\cite{bernsten1991}. 
%
%

In Fig.~\ref{fig:gapped}~(c) the results for the static polarizability $\Pi(\qq,\omega=0)$ are shown
for fixed doping $n=10^{12}~{\rm cm}^{-2}$ and different values of $\Delta$.
We see that for $\Delta\neq 0$ the results obtained with the 4-band model are very different
from the ones obtained with the 2-band model. In the 2-band model $\Pi(\qq,0)$ exhibits a Kohn anomaly
only for $q=2k_F$ ($k_F$ being the Fermi wavevector), whereas in the 4-band model (and the refined 2-band model)
$\Pi(\qq,0)$ exhibits Kohn anomalies also for values of $q<2k_F$.
This is due to the fact that in the 4-band model, at low energies, the lowest bands, for $\Delta\neq 0$,
acquire a typical nonmonotonic sombrero shape. As a consequence in the 4-band model,  for $\Delta\neq 0$,
for fixed $n$ ($\Delta$) when $\Delta>\Delta_c\df\hbar v_F\sqrt{\pi n}$
($|n|<n_c\df\Delta^2/(\pi\hbar^2 v_F^2)$)
the Fermi surface is multiply connected.
Neglecting trigonal warping for $n<n_c$ the Fermi surface is formed
by two circumferences, of radii $k_{F\pm}=(1/\hbar v_F)\sqrt{\eps_F^2+\Delta^2/4\pm\sqrt{\eps_F^2(\gamma_1^2+\Delta^2)-\Delta^2/4}}$ respectively,
with $\eps_F=(1/2)\sqrt{(\hbar^4v_F^4\pi^2n^2+\Delta^2\gamma_1^2)/(\gamma_1^2+\Delta^2})$
(see inset of Fig.~\ref{fig:gapped}~(b)). In this situation we can expect additional Kohn anomalies for values of $q$ joining points on the same connected part of the Fermi surface
and on disconnected parts of the Fermi surface.
For $\gamma_1=350$~meV and $n=10^{12}~{\rm cm}^{-2}$ we have that $\Delta_c\approx\gamma_1/3$.
For $\Delta=\Delta_c$ the Fermi energy
just touches the top of the sombrero. In this case we only have one additional Kohn anomaly
for $q=k_F$ in addition to the $q=2k_F$ one, see Fig.~\ref{fig:gapped}~(c).
For $\Delta>\Delta_c$
the Fermi energy cuts the sombrero region and so
we have Kohn anomalies
for $q=k_{F+}-k_{F-}$, and $q=2k_{F-}$ in addition to the one for $q=2k_{F+}$
as shown in  Fig.~\ref{fig:gapped}~(c).
One might expect to observe an anomaly also
for $q=k_{F+}+k_{F-}$, however the points on the Fermi surface connected by this value
of $q$ have Fermi velocities with the same sign and therefore the anomaly is suppressed.
Fig.~\ref{fig:gapped}~(d) shows the dependence of $\Pi(\qq,0)$
on $q$ and $n$ for $\Delta=\gamma_1/2$. From this figure we see the evolution of the Kohn anomalies with doping,
in particular we can observe the merging of some of the anomalies for specific
values of the doping.

\begin{figure}[htb]
 \begin{center}
  \centering
  \includegraphics[width=8.5cm]{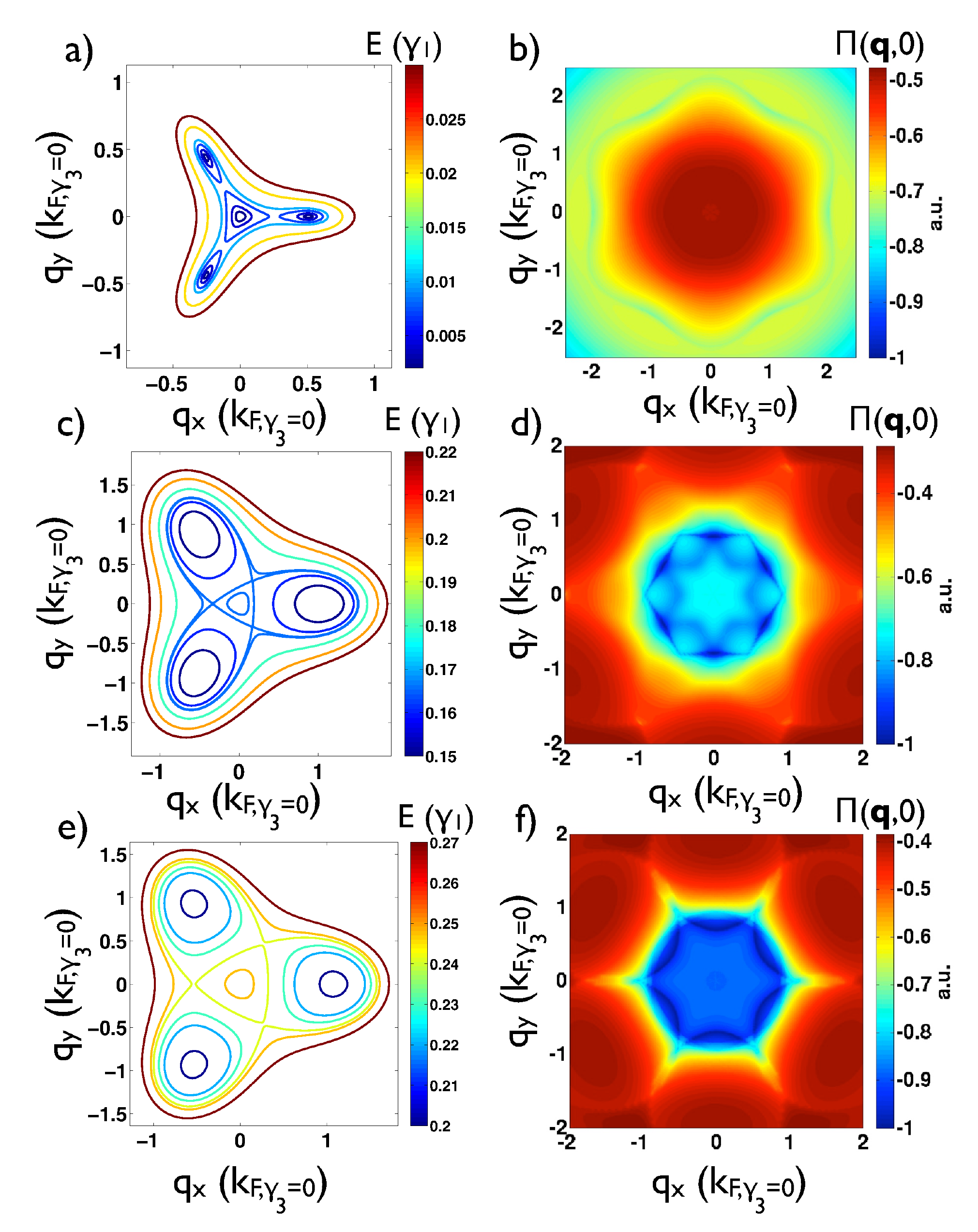}
  \caption{
           (Color online).
           (Left Column) Equipotential lines for the lowest energy band within the 4-band model with $\gamma_3=3\gamma_1/4$
           and $\Delta=0,\ 1/3\gamma_1,\ 1/2\gamma_1$
           from top to bottom.
           (Right Column) $\Pi(\qq,0)$ for $n=10^{12}~{\rm cm}^{-2}$, trigonal warping $\gamma_3=3\gamma_1/4$, and $\Delta=0,\ 1/3\gamma_1,\ 1/2\gamma_1$
           from the top panel to bottom one. $k_{F,\gamma_3=0}$ is $k_{F+}$ in the limit $\gamma_3=0$.
          }
  \label{fig:gapped_trig}
 \end{center}
\end{figure}

We now consider the effects on $\Pi(\qq,0)$ of trigonal warping.
As shown by the left panels of Fig.~\ref{fig:gapped_trig}, in the presence of trigonal warping the energy-bands
become anisotropic
\cite{mccann2006b, nilsson2008}. In particular, at low energies the lowest bands exhibit 4 degenerate
minima. The modifications of the fermionic energy bands due to the trigonal warping are reflected in the polarizability, as
shown by the right panels of  Fig.~\ref{fig:gapped_trig}. $\Pi(\qq,0)$ becomes strongly anisotropic,
the number and position of the Kohn anomalies becomes dependent on the direction of $\qq$.

\begin{figure}[htb]
 \begin{center}
  \centering
  \includegraphics[width=8.5cm]{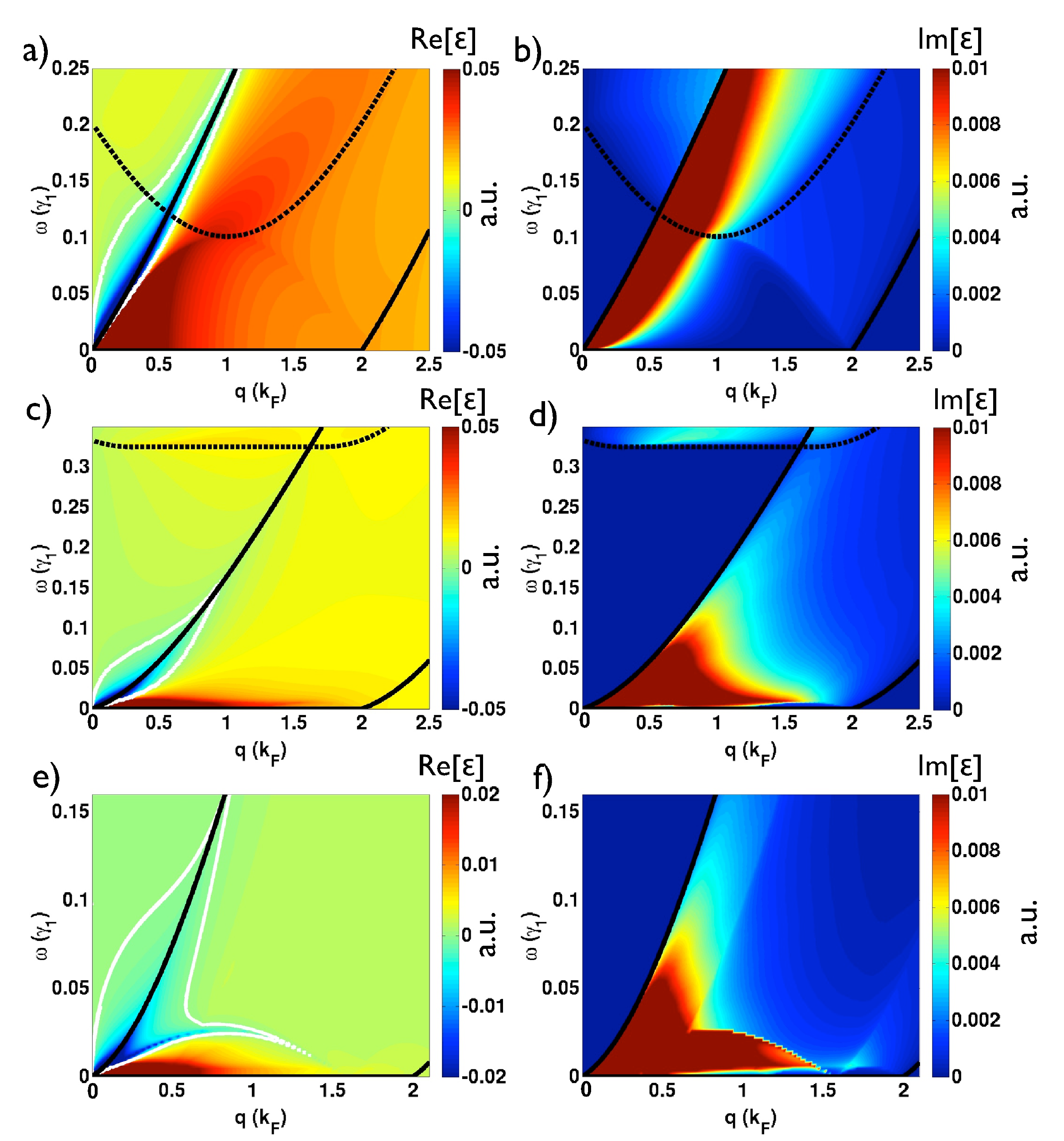}
  \caption{
           (Color online). The left (right) column shows the real (imaginary) part of $\epsilon_{\text{RPA}}(\textbf{q},\omega)$
           for $\Delta=0$, $\Delta=\gamma_1/3$, $\Delta=\gamma_1/2$ from top to bottom.
           The plasmon dispersion is denoted by white curves. The boundaries for the intraband (interband) particle-hole continuum are
           indicated with black solid (dashed) curves.
          }
  \label{fig:epsilon_alpha0.5}
 \end{center}
\end{figure}

The dynamic dielectric function \dfun for fixed doping $n=10^{12}~{\rm cm}^{-2}$
and $\Delta<\Delta_c$, $\Delta=\Delta_c$, $\Delta>\Delta_c$ for the case in which
$\gamma_3=0$ (no trigonal warping) is shown in Fig.~\ref{fig:epsilon_alpha0.5}.
The white lines show the plasmon dispersion, the black solid (dashed) lines
show the boundaries of the intraband (interband) particle-hole continuum.
We see that as $\Delta$ crosses $\Delta_c$ the dispersion of the plasmon mode
outside the particle-hole continuum is not modified qualitatively.
The plasmon mode inside the particle hole-continuum on the other hand is qualitatively
very different for $\Delta<\Delta_c$ and $\Delta>\Delta_c$, an effect that is not
captured by the 2-band model \cite{wang2010}.

\begin{figure}[htb]
 \begin{center}
  \centering
  \includegraphics[width=8.5cm]{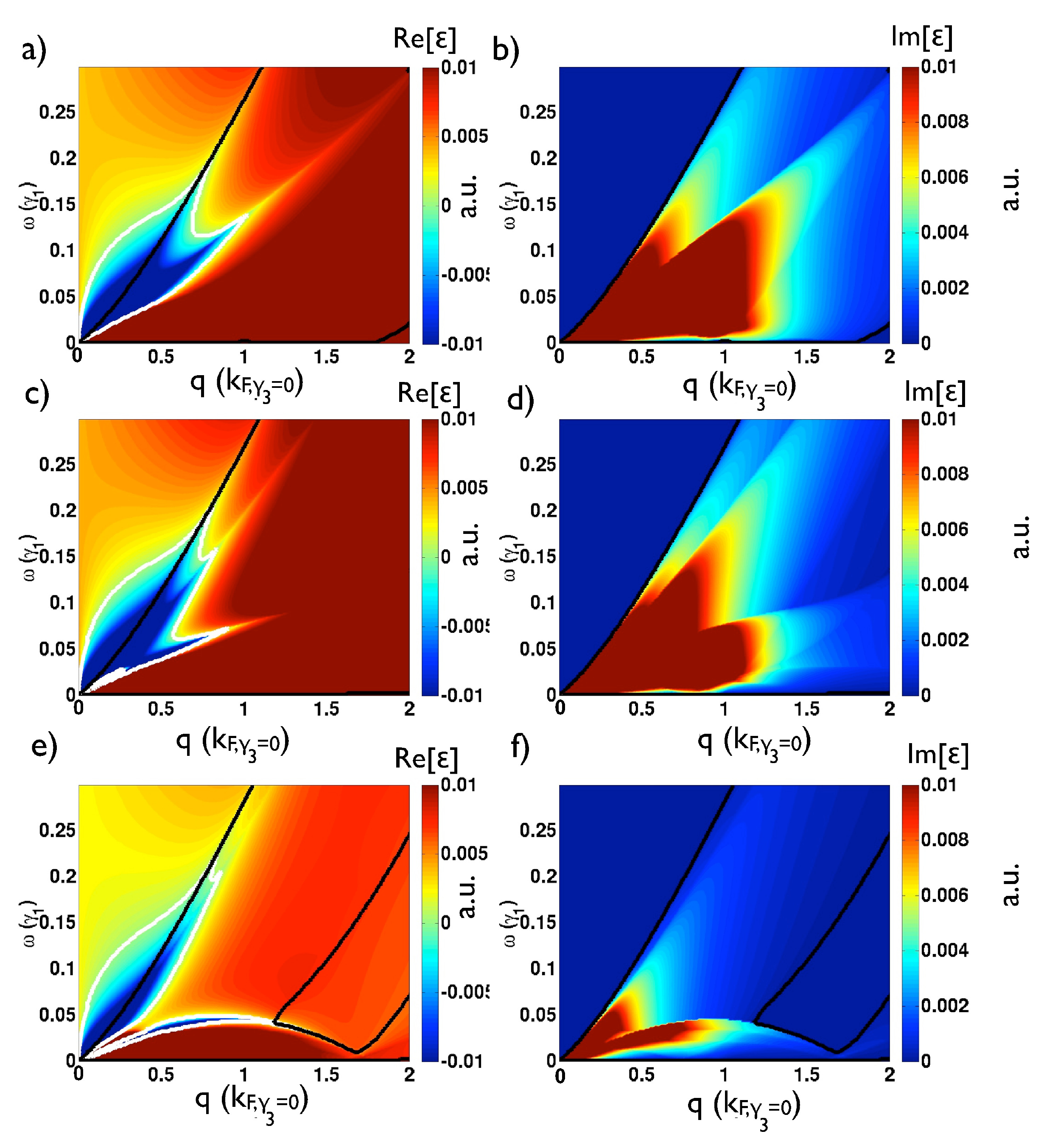}
  \caption{
           (Color online). 
           The left (right) column shows the real (imaginary)
           part of $\epsilon_{\text{RPA}}(\textbf{q},\omega)$
           with $\Delta=\gamma_1/2$ for $\theta=0^\circ,\ 15^\circ, 30^\circ$.
           The white curves denote the plasmon dispersion, the black curves denote the boundaries of the particle-hole continuum.
           $k_{F,\gamma_3=0}$ is $k_{F+}$ in the limit $\gamma_3=0$.
          }
  \label{fig:epsilon_trig}
 \end{center}
\end{figure}

In the presence of trigonal warping \dfun becomes strongly anisotropic and this is particularly
evident when the Fermi energy cuts the sombrero region.
Fig.~\ref{fig:epsilon_trig} shows the results for \dfun for different directions of $\qq$
obtained taking into account trigonal warping.
From the figure the strong
anisotropy of \dfun when $\gamma_3\neq 0$ is evident. In particular, we see that the plasmon mode
inside the p-h continuum is very different for different directions of $\qq$.
%

\begin{figure}[htb]
 \begin{center}
  \centering
  \includegraphics[width=8.5cm]{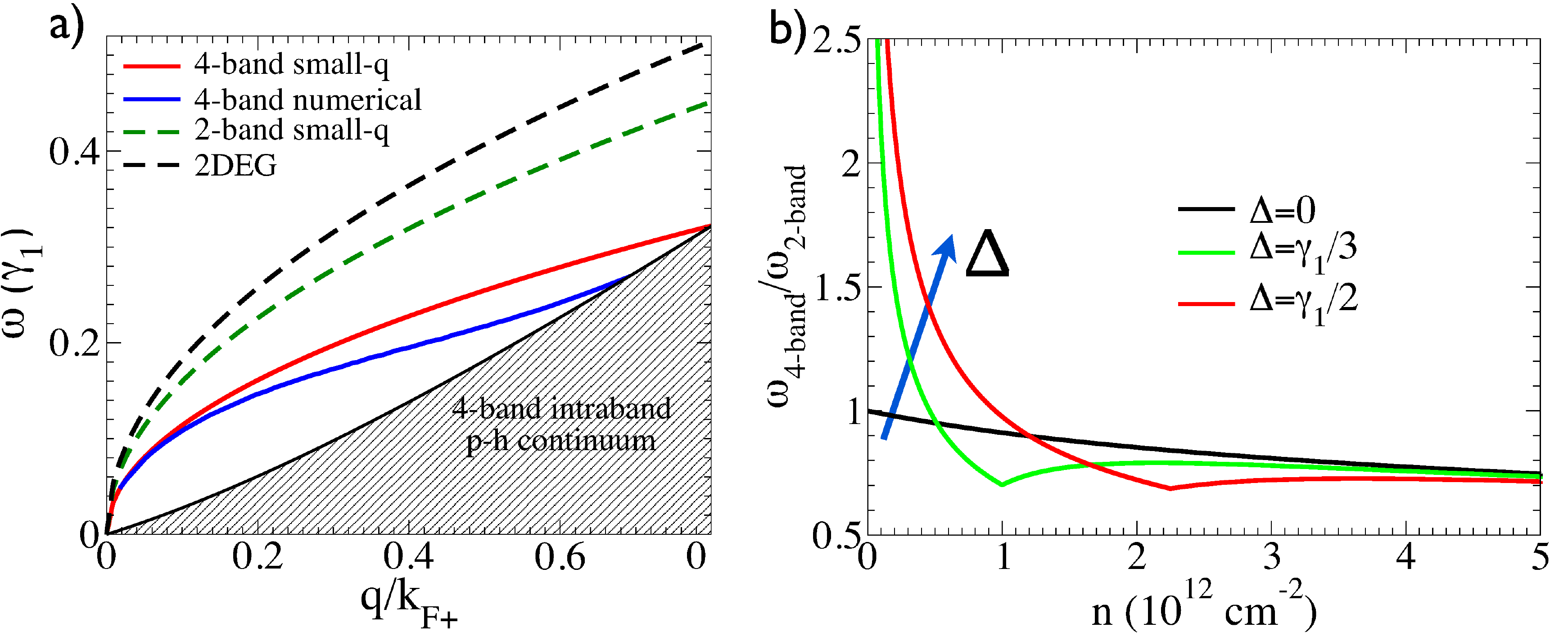}
  \caption{
           (Color online)
           (a) 
           Plasmon dispersion for $\Delta=\gamma_1/2$ obtained using the full 4-band model,
           and the 2-band model for $n=2.7\times 10^{12}{\rm cm^{-2}}$.
           (b) Ratio $\omega_{{\rm 4-band}}/\omega_{{\rm 2-band}}$ as a function of doping for different value of $\Delta$.
           For $\Delta\neq 0$ and $n\to 0$ the ratio $\omega_{{\rm 4-band}}/\omega_{{\rm 2-band}}$ diverges.
          }
  \label{fig:plasmon}
 \end{center}
\end{figure}

For the case with no trigonal warping
in the long-wavelength limit $q\ll\omega/v_F$ using the 4-band model
for the polarizability, up to order $q^2$, we have:
\beq
 \Pi(\qq,\omega)=\frac{g q^2}{4\pi\omega^2}
                \left[
                  k_{F+}\left.\frac{\partial\eps_{\kk}}{\partial\kk}\right|_{k_{F+}} -
                  k_{F-}\left.\frac{\partial\eps_{\kk}}{\partial\kk}\right|_{k_{F-}}
                \right]
 \label{eq:pi}
\enq
We notice that in Eq.~\ceq{eq:pi} there is a term proportional to $k_{F-}$ that is
absent in the 2-band model.
Replacing this expression in the equation for the RPA \dfun we find the plasmon
dispersion:
\beq
 \omega = \left[\frac{g}{2}\hbar v_F\alpha q
          \left(
                  k_{F+}\left.\frac{\partial\eps_{\kk}}{\partial\kk}\right|_{k_{F+}} -
                  k_{F-}\left.\frac{\partial\eps_{\kk}}{\partial\kk}\right|_{k_{F-}}
                \right)
          \right]^{1/2}.
 \label{eq:plasmon}
\enq
This dispersion is very general and is valid both for $n<n_c$ and $n>n_c$,
in the latter case $k_{F-}=0$.
From Eq.~\ceq{eq:plasmon} using 
the appropriate expressions for $k_{F+}$, $k_{F-}$  and $\eps_\kk$ we find 
\beq
 \omega(q)=\sqrt{\dfrac{qge^2\gamma_1^2}{\epsilon_F\epsilon}}F(\hat n,\hat\Delta)
 \label{eq:plasmon2}
\enq
where
$\hat n\equiv\hbar^2v_F^2\pi n/\gamma_1^2$, 
$\hat\Delta\equiv\Delta/\gamma_1$,
and, for the 2-band model, 
$\eps_F=\gamma_1[\hat{n}^2+\hat{\Delta}^2/4]^{1/2}$,
$F(\hat n,\hat\Delta)=\hat n$, whereas
for the 4-band model in the sombrero region,
$\eps_F=\gamma_1[(\hat{n}^2+\hat{\Delta}^2)/(1+\hat{\Delta}^2)]^{1/2}/2$,
$F(\hat n,\hat \Delta)=[\hat n (\hat{\Delta}^4+2\hat{\Delta}^2-\hat n^2)/(\hat{\Delta}^4+2\hat{\Delta}^2-\hat n^2+1 )]^{1/2}/2$,
and outside the sombrero region,
$\eps_F=\gamma_1[2+\hat{\Delta}^2+4\hat{n}-2[1+4\hat{n}(1+\hat{\Delta}^2)]^{1/2}]^{1/2}/2$,
$F(\hat n,\hat\Delta)=[(\hat n/2)[1-(1+\hat{\Delta}^2)]/[1+4\hat n(1+\hat{\Delta}^2)]^{1/2}]^{1/2}$.
In Fig.~\ref{fig:plasmon}~(a) we compare the results for
the plasmon dispersion obtained numerically using the 4-band model with the ones given by Eq.~\ceq{eq:plasmon2}
for the gapped case $\Delta=\gamma_1/2$ for a given value of $n$. 
We see that the 2-band results differ substantially from the 4-band results.
At low densities ($n<n_c$) this is due to the fact that the 2-band model does not
capture the nonmonotonic band structure, i.e. the fact that in the 2-band model
in Eq.~\ceq{eq:plasmon} there is no term $k_{F-}\partial\eps_{\kk}/\partial\kk|_{k_{F-}}$.
For $n>n_c$ this is due to the fact that in the 4-band model the dispersion is much
closer to linear than parabolic as assumed in the 2-band model, in analogy to what happens
in the gapless case \cite{sensarma2010}. As a consequence 
for very large $n$ we have $\omega_{{\rm 4-band}}/\omega_{{\rm 2-band}}\propto n^{-1/4}$.
This is summarized in  Fig.~\ref{fig:plasmon}~(b)  which shows the ratio $\omega_{{\rm 4-band}}/\omega_{{\rm 2-band}}$
between the plasmon frequency obtained within the 4-band and the 2-band model as a function of $n$
for different values of $\Delta$. Notice that in the long-wavelength limit this ratio (see Eq.~\ceq{eq:plasmon2})
is independent of $q$ and is a function only of $n$ and $\Delta$.
%

In conclusion, we have obtained the static and dielectric screening of gapped BLG
using the full 4-band model. We find that the static screening obtained using
the 4-band model is qualitatively different from the one obtained from the 2-band model.
In particular in the 4-band model, when the gap is nonzero,
the static polarizability exhibits Kohn anomalies that are not present in
the simplified 2-band model.
For the dynamic screening we have found that the plasmon frequency within the 
4-band model is substantially different from the one obtained within the 
2-band model especially at low densities when $\Delta\neq 0$.
%
We have also characterized
the strong anisotropic properties of the static and dynamic screening due to the
trigonal warping. We find that in the presence of trigonal warping
in gapped BLG the number of Kohn anomalies depends not only on the doping and the band-gap but
also on the direction of the momentum.
Our results, in particular the identification of additional Kohn anomalies, and the
strong anisotropic nature of the screening in the presence of trigonal warping,
have important implications for understanding of
the phonon spectrum and the nature of the RKKY interaction in gapped BLG,
and are therefore expected to have clear experimental signatures.
Moreover our results apply also to the case in which a gap opens due to
the realization of a spontaneously broken symmetry state and could
then be used to identify and characterize such a state.

This work is supported in part by the Jeffress Memorial Trust, Grant No. J-1033, and by a
faculty research grant from the College of William \& Mary.
CT acknowledges support from the Virginia Space Grant Consortium.
Computations were carried out on the SciClone Cluster at the College of William \& Mary.

%


\begin{thebibliography}{41}
\bibitem[{not()}]{note01}
\bibinfo{note}{The closeness of the two layers induces the finite interlayer
  hopping, this implies that BLG has a band structure different from the one of
  two independent graphene layers.}
\end{thebibliography}



\end{document}